\documentclass[prb,twocolumn,showpacs]{revtex4}
\usepackage{graphicx}
\usepackage{bm}
\usepackage{amssymb,amsmath}
\usepackage[usenames]{color}
\usepackage{epsfig}
\usepackage{hyperref}
\hypersetup{colorlinks=true, citecolor=blue,
linkcolor=blue,urlcolor=blue }

\begin{document}

\title{Universality of Anderson transition in two-dimensional systems of symplectic symmetry class}
\author{Reza Sepehrinia}
\address{School of Physics, Institute for Research in Fundamental Sciences, IPM, 19395-5531 Tehran, Iran}
\begin{abstract}
We investigate localization of noninteracting particles with spins
higher than $\frac{1}{2}$ in a two-dimensional random potential in
presence of spin-orbit coupling. We consider an integer spin ($s=1$)
and a half-integer spin ($s=\frac{3}{2}$) belonging to orthogonal
and symplectic symmetry classes, respectively. We show that
particles with integer spin are localized and those with
half-integer spin exhibit Anderson transition. The transition
belongs to universality class of conventional symplectic model for
spin-$\frac{1}{2}$ particles.
\end{abstract}
 \maketitle
\section{Introduction}
Symplectic class has a rich physical content among the symmetry
classes in the Wigner-Dyson classification of random matrices.
Spin-orbit scattering provides the common physical realization of
this class. From the symmetry point of view such system is invariant
under time reversal ($\mathcal{T}$) but not under spin rotation
($\mathcal{S}$). For a system with non-integer spin we have
$\mathcal{T}^2=-1$ and the wave function (WF) has a rotational
periodicity of $4\pi$ therefore in average, time-reversed paths in
the multiple-scattering picture interfere
destructively.\cite{Bergmann82} As a result of the destructive
interference, conductivity is enhanced and we have weak
anti-localization rather than weak localization. This is one of the
mechanisms of criticality in two spatial dimensions which is
specific for symplectic class in the framework of early Wigner-Dyson
classification.

Several other realizations of symplectic class have been identified
which exhibit distinct universal behavior. For this class the
homotopy group of $\sigma$ model manifold is nontrivial so the
$\sigma$ model action allows for inclusion of a topological $\theta$
term (see Ref. \onlinecite{Evers08} for a recent review). Such
topological term is responsible for quantum Hall criticality in the
unitary class. Spin-orbit coupling have provided similar topological
phases in the $\mathcal{T}$-invariant systems which has been the
subject of an intense activity in recent years.\cite{Topol}
Remarkably, quantum spin Hall (QSH) which is a novel phase induced
by $Z_2$ topological term in the symplectic ensemble. A new
universality class of Anderson transition emerges in presence of
this topological structure between the metallic and QSH
phases.\cite{Onoda07} It should be mentioned however that the
corresponding Chalker-Coddington network model which allows to have
odd number of Kramers doublets (i.e., nontrivial topology) does not
capture this critical behavior.\cite{Obuse07} Another realization of
symplectic class with nontrivial topology appears in a
two-dimensional system of Dirac fermions
\cite{Ostrovsky07,Nomura-Bardarson} which yields a unconventional
scaling $\beta$ function. Two different scenarios are proposed one
of which predicts an extra attractive fixed point in the
strong-coupling limit\cite{Ostrovsky07} and the other one implies on
delocalization of all states even in strong disorder
limit.\cite{Nomura-Bardarson} Based on semiclassical arguments it is
also shown that spin-orbit scattering may induce a novel
universality class in the regime of integer quantum-Hall
effect.\cite{Avishai02} Whereas in the unitary ensemble, presence or
absence of spin-rotational invariance does not change symmetry
class. These examples imply the fact that in spite of a complete
mathematical classification of symmetry classes, universality
classes are not recognized so far.

Appreciate to new advances in designing periodic potentials by
standing waves of light, many experiments which are not possible to
arrange for electrons can be simulated with quantum motion of cold
atoms. Especially it just recently became possible to do careful
experiments on localization of noninteracting \textit{matter} waves.
Even more fascinating games could be done by changing the
polarization of the beams. So the internal degrees of freedom of the
atom can be coupled to the momentum of the beam and produce an
effective spin-orbit like term in the Hamiltonian. Once this could
be done, one can search for new universalities and topological
properties of WFs with higher tunability.\cite{Dudarev04}

A natural generalization in this direction is to consider atoms with
higher number of internal degrees of freedom or particles with
higher spins. Here we want to address whether higher spins in
spin-orbit interaction can change the universality class of
transition or not. We use the transfer-matrix method to calculate
the localization length and then extract the critical exponents from
finite-size scaling analysis. On the other hand
\textit{multifractal} spectrum of critical WFs are of universal
properties hence useful to describe the transition. We will examine
multifractal properties of higher spin model in comparison with
spin-$\frac{1}{2}$ case.
\section{Model}
Several models have been proposed to study Anderson localization
problem in presence of spin-orbit scattering.\cite{Ando89}
Regardless of microscopic details, they present the same universal
features. Spin-relaxation length is an important irrelevant length
scale in these systems. SU(2) model\cite{Asada02} has smaller
spin-relaxation length since spin rotation operators in each link of
its lattice, are uniformly distributed. As a result it needs small
correction to scaling. Having small size effects this model provides
more accurate calculation of critical exponents. Here we use the
generalization of this model to describe particles with higher
spins. We start with the following Hamiltonian which is proposed for
spin-$\frac{1}{2}$ particles.
\begin{equation}\label{Hamiltonian}
H=\sum_{i \sigma} \epsilon_i c^{\dagger}_{i \sigma}
c_{i\sigma}-V\sum_{\langle ij \rangle \sigma \sigma'}
\mathcal{R}^{ij}_{\sigma \sigma'}c^{\dagger}_{i \sigma} c_{j\sigma'}
\end{equation}
Here we will let the hopping matrices act on spinors of higher rank.
So spin indexes ($\sigma,\sigma'$) take the values ($s, s-1, \cdots
,-s$) for particles with spin $s$. Latin indices denote
nearest-neighbor sites on square lattice. Random on-site potential
$\epsilon_i$ is distributed uniformly in the interval
$[-\frac{W}{2},\frac{W}{2}]$. Energy scale will be set by $V=1$.

$\mathcal{R}^{ij}$'s are ($2s+1$)-dimensional irreducible
representation of SU(2) group. In terms of Euler angles they have
the following description $\mathcal{R}^{ij}_{\sigma
\sigma'}=D^s_{\sigma
\sigma'}(\alpha_{ij},\beta_{ij},\gamma_{ij})=e^{i(\sigma
\alpha_{ij}+\sigma' \gamma_{ij})}d^s_{\sigma \sigma'}(\beta_{ij})$,
where $d^s_{\sigma \sigma'}(\beta_{ij})=\langle s
\sigma|e^{-i\beta_{ij}S_y/\hbar}|s \sigma'\rangle$ is the matrix
element of rotation operator around $y$ axis. Angles $\alpha$,
$\beta$, and $\gamma$ are distributed randomly in different links of
lattice such that rotation matrices $\mathcal{R}^{ij}$ have uniform
distribution with respect to the Haar measure on SU(2) group.
Namely, $\alpha$ and $\gamma$ are distributed uniformly in the
interval $[0,2\pi)$ and $\beta$ is chosen from interval
$[0,\frac{\pi}{2}]$ with distribution $P(\beta)=\sin(2\beta)$.
\section{Level Statistics}
Regarding broken spin-rotational symmetry of Hamiltonian
(\ref{Hamiltonian}), for integer spins it does belong to the
orthogonal ensemble. Only for half-integer spins it falls into the
symplectic ensemble. This is known from Wigner-Dyson classification
of random matrices.\cite{Mehta} An essential difference in the
spectrum of two cases is Kramers degeneracy of energy levels of
later which is robust against disorder.

\begin{figure}[t]
\epsfxsize7truecm \epsffile{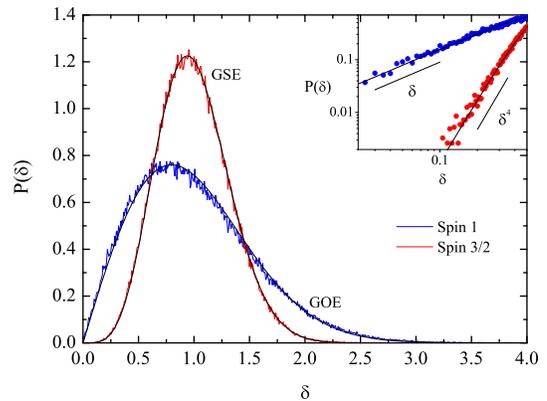} \caption{(Color online)
Level-spacing distribution function for disorder strength $W=0$.
Spin 1(GOE), spin $\frac{3}{2}$(GSE) and corresponding Wigne-Dyson
distributions with $\beta=1,4$, respectively. Inset: log-log plot
which shows power-law behavior at $\delta \rightarrow 0$. Line
segments represent functions proportional to $\delta$ and $\delta^4$
.}\label{LevelStat}
\end{figure}

In this section we demonstrate above-mentioned relation with the
Wigner-Dyson symmetry classes numerically. The simplest quantity
which can be used to determine the statistical properties of energy
levels is the nearest level-spacing distribution. To have comparable
results with the distributions of random matrix theory we look at
the distribution of $\delta_n=\frac{E_{n+1}-E_n}{\langle E_{n+1}-E_n
\rangle}$ which is unfolded level spacing. Denominator is the
ensemble average of level spacing which is proportional to inverse
density of states (DOS). So unfolding procedure is needed when the
DOS has large variations within the energy range under
consideration. Obviously first-order moment of distribution function
$P(\delta)$ is fixed, i.e., $\langle \delta\rangle=\int_0^{\infty}
\delta \ P(\delta) d\delta=1$ for unfolded spectrum. We determine
the distribution function $P(\delta)$ for two $s=1$ and
$s=\frac{3}{2}$ cases in the metallic regime, after removing the
degeneracy in the later case. By metallic regime we mean weak
disorder for which WFs have large overlap and comparable
localization length with the system size. We find good agreement
with GOE ($\beta=1$) for $s=1$ and GSE ($\beta=4$) for
$s=\frac{3}{2}$ cases. The results which are shown in Fig.
\ref{LevelStat} are obtained by diagonalizing $10^3$ Hamiltonians of
lattice size $20^2$ and disorder width $W=0$. For nonzero but small
values of $W$ also we obtain the same results. In the strong
disorder limit which all states (for both cases) tend to be
localized one naturally expect to see Poisson distribution.
Difference between two cases would reveal in the intermediate
disorder strength.
\section{Transfer Matrix}
To find a precise insight in to the localization properties of these
models and to explain the differences in the thermodynamic limit we
implement a finite-size scaling analysis. In the following we will
study renormalized localization length (RLL),
$\Lambda=\frac{\lambda_{_m}}{M}$, on the quasi-one-dimensional
geometry, where $\lambda_{_m}$ is the localization length on strip.
We utilize transfer-matrix method \cite{Markos06} to calculate the
minimal Lyapunov exponent, inverse of which is the largest length
scale of spatial extension of wave function. Dimension of transfer
matrices for spin $s$ is $N=2(2s+1)M$ with $M$ being the system size
in the transverse direction. Lyapunov exponents appear in
$(-\gamma,\gamma)$ pairs for integer spin case due to symmetry of
transfer matrices. Furthermore for half-integer spin case each
$\gamma$ appears twice due to Kramers degeneracy. According to these
symmetries we need to evolve $\frac{N}{2}=(2s+1)M$ vectors of length
$N$ to calculate minimum positive $\gamma$. Components of vectors
are $V_{(2s+1)j+m}=\psi^{n,j}_{\sigma},
V_{(2s+1)(j+M)+m}=\psi^{n-1,j}_{\sigma}$, where $n$ is the number of
layer (here a chain of length $M$), $j=0,\ldots,M-1$ is the
coordinate in the transverse direction and $m$ takes values
$1,\ldots,(2s+1)$ corresponding to $\sigma=s, s-1,\ldots,-s$,
respectively. Gram-Schmidt orthogonalization is implemented after
each four steps.

\begin{figure}[t]
\epsfxsize8truecm \epsffile{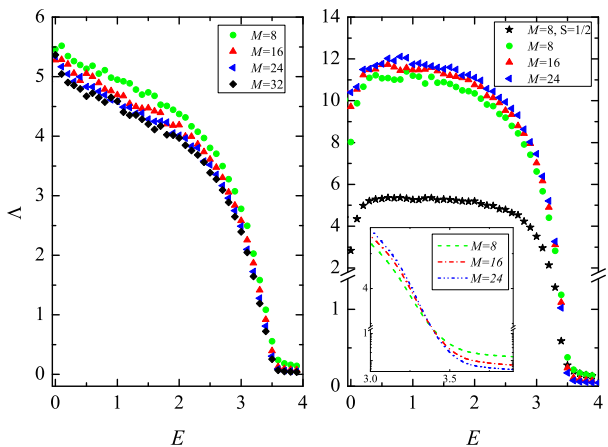} \caption{(Color online) Left:
RLL as a function of energy of spin-1 particle with zero on-site
disorder $(W=0)$, right: spin $\frac{3}{2}$ and spin $\frac{1}{2}$
(for single size $M=8$) with $W=1$. Inset shows a zoom in around the
crossing point.}\label{TM}
\end{figure}

Let us start with spin-$1$ particle and zero on-site disorder
($W=0$). In the left panel of Fig. \ref{TM} we observe that
$\Lambda$ decreases by increasing the size $M$ in the whole energy
range. We can conclude that states are localized even for zero
on-site disorder. In other words the randomness in spin rotation in
passing through different links is enough to localize the particle.
Results for nonzero on-site disorder are the same and we will not
present them here. This is what we expect for a system in orthogonal
(AI) symmetry class. That breaking of spin-rotational symmetry in a
$\mathcal{T}$-invariant system with integer spin neither changes
symmetry class nor develops delocalized states. We should comment on
the additional symmetry which Hamiltonian (\ref{Hamiltonian}) may
have in absence of on-site disorder. Using periodic boundary
conditions in the transverse direction and even $M$ the lattice will
be a \textit{bipartite} lattice which is shown to have anomalies at
zero energy.\cite{Miller96,Brouwer98,Brouwer00} We can see in Fig.
\ref{TM} that at the center of energy band, $\Lambda$ remains almost
constant for different sizes which indicates a critical state at
$E=0$. Away from the band center $\Lambda$ decreases more rapidly by
increasing $M$, which gradually leads in creation of a cusp at
$E=0$. For an odd number of channels ($M$) with free boundary
conditions in transverse direction particularly
$\Lambda\rightarrow\infty$ for this zero mode. This is
proved\cite{Brouwer98} analytically for coupled one-dimensional
chains with $\beta=1,2$. This critical state will be ruined by
addition of small on-site disorder which breaks sublattice symmetry
of Hamiltonian.
Unlike the spin-1 particle the case with spin-$\frac{3}{2}$
possesses a band of extended states for certain values of on-site
disorder strength $W$. As an example, results of $\Lambda$ for
spin-$\frac{3}{2}$ particle with disorder strength $W=1$ and sizes
$M=8,16,24$ are shown in Fig. \ref{TM}. The result for
spin-$\frac{1}{2}$ particle is also included for comparison at the
same on-site disorder strength. Relative errors of data are $1\%$
for $M>8$, $0.8\%$ for $M=8$, and $0.7\%$ for spin-$\frac{1}{2}$
case. To reach this accuracy the length of strips is increased up to
$1.3, 1.6, 2.4, 3.6\times 10^{6}$ for spin $1$ with $M=8, 16, 24,
32$, $1.4, 1.8, 2.8, 3.8\times 10^{6}$ for spin $\frac{3}{2}$ with
$M=8, 16, 24, 32$, respectively, and $0.9\times 10^{6}$ for
spin-$\frac{1}{2}$ case. There can be seen a mobility edge at
$E\approx3.4$ where order of symbols is reversed. As usual the
states in the band edge are localized and the mid-band states are
extended. The finite band of the bulk extended states make the true
metallic phase happen in the symplectic class. By increasing the
disorder strength the band of extended states gets narrower and
gradually collapses at a critical value $W_c$.

\begin{figure}[t]
\epsfxsize7.5truecm \epsffile{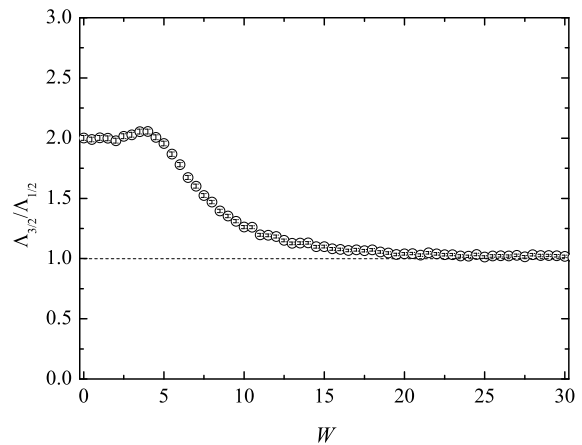} \caption{Ratio of RLLs vs
$W$ of spin-$\frac{3}{2}$ and spin-$\frac{1}{2}$ particles for
single energy $E=2$ and $M=8$.}\label{ratio}
\end{figure}

It is worthwhile to compare RLL of spin-$\frac{3}{2}$ and
spin-$\frac{1}{2}$ cases more closely. The ratio of RLLs vs $W$ of
two models is plotted in Fig. \ref{ratio} for single energy $E=2$
and $M=8$. The energy is chosen away from the band center and edges
to ensure the DOS has considerable value in both models. In a range
of weak disorder strengths ($W\lesssim4$), the ratio is nearly
constant and equals 2. This is where the hopping term is dominant or
comparable with on-site term. In strong disorder limit ($W\gg1$) the
hopping term is negligible, therefore spin degrees of freedom would
not have considerable effect on localization length. Thus one
expects the same RLL for both cases. That is what which can be seen
also in Fig. \ref{ratio} at large $W$.
In the next sections we will discuss critical exponents
characterizing the universality class of the transition.
\section{Scaling and Critical Indices}
Dimensionless quantity $\Lambda$ is one of scaling variables which
is frequently used for numerical analysis. On the basis of
one-parameter scaling hypothesis, it can be written in the following
form:
\begin{equation}\label{Lambda}
    \Lambda(E,W,M)=f\left(\frac{M}{\xi(W,E)}\right)
\end{equation}
where $\xi(W,E)$ is the localization length (insulating side) or the
correlation length (metallic side) of infinite system. It is not the
only length scale in this system. We will encounter deviations from
scaling, Eq. (\ref{Lambda}), when other (irrelevant) length scales
are comparable with correlation length. Near the mobility edge,
$\xi$ diverges as $\xi\sim |E-E_c|^{-\nu}$ with critical exponent
$\nu$. The value of $\Lambda$ at critical point is also a universal
constant and independent of $W$.

To calculate the critical exponent we take few sets of data
($\Lambda$ vs $E$) close to the mobility edge for $M=8,16,24,32$.
There are two ways of fitting the data to scaling form
(\ref{Lambda}). Since function $f(x)$ is unknown we can either use a
Taylor expansion and then obtain the coefficients by fitting, or
take one set as $f(x)$ (by interpolating), define a suitable
residual of curves and minimize it by adjusting critical indices.
Let the values of $\Lambda$ and $E$ in the $i$th set (corresponding
to size $M_i$) be denoted by $\Lambda_{ij}$ and $E_{ij}$.
\begin{figure}[h]
\epsfxsize7truecm \epsffile{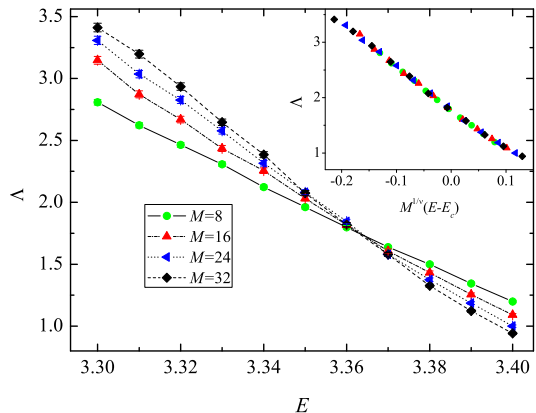} \caption{(Color online) RLL of
spin-$\frac{3}{2}$ particle for $W=1$ and sizes $M=8,16,24,32$.
Inset: after rescaling.}\label{FSS}
\end{figure}

A possible definition of residual follows\cite{Bhattacharjee01}
\begin{equation}\label{Residual}
    R=\frac{1}{N}\sum_k \sum_{i\neq k} \sum_j{}^{{}'}|\Lambda_{ij}-\Lambda_k(M^{1/\nu}(E_{ij}-E_c))|
\end{equation}
where $\Lambda_k(x)$ is obtained by transforming the horizontal axis
of $\Lambda_{kj}$ such as $E_{kj}\rightarrow
M_k^{1/\nu}(E_{kj}-E_c)$ and simple linear interpolation. The prime
on the third sum denotes summation over $j$'s which are in the range
of definition of $\Lambda_k(x)$ and $N$ is the total number of such
points. For each $k$ one set is taken as the reference curve and the
other curves are supposed to collapse on it by rescaling. Function
$R$ reaches its minimum value ($\sim0.01$) at $\nu=2.81\pm 0.18$ and
$E_c=3.362\pm 0.014$. Error-bars are roughly estimated from width of
minimum by using approximate expressions given in Ref.
\onlinecite{Bhattacharjee01}. Reasonable data collapse is obtained
for these values of critical parameters (inset of Fig. \ref{FSS}).

We tabulate the values of critical exponents in Table I. The
exponents of spin-$\frac{1}{2}$ system are also given for
comparison. In spite of relatively larger error-bars in the
spin-$\frac{3}{2}$ case, estimated exponent is close to that of
spin-$\frac{1}{2}$ particle. Our speculation is that two models
belong the same universality class.
\begin{table}[h]
\caption{Critical exponents: results of spin-$\frac{1}{2}$ case are
taken from Ref. \onlinecite{Asada02}.} \centering
\begin{tabular*}{0.48\textwidth}{@{\extracolsep{\fill}}c c c}
\hline\hline Model & $\nu$ & $\Lambda_c$ \\ [0.5ex] \hline
Spin $\frac{1}{2}$ & $2.73 \pm 0.02$ & $1.844 \pm 0.001$  \\
Spin $\frac{3}{2}$ & $2.81\pm 0.18$ & $1.77 \pm 0.07$  \\
 [1ex]
\hline
\end{tabular*}
\label{TableI}
\end{table}
\section{Multifractal Spectrum}
Fluctuations of critical WFs exhibit universal features. So
multifractal exponents which describe the distribution of WFs can be
used to characterize phase transition. Moreover multifractality
leads to anomalous diffusion of wave packets near the mobility edge.
This is reflected in power-law decay of return probability which is
governed by one of those multifractal exponents.  We want to compare
the multifractal spectrum of spin-$\frac{3}{2}$ and
spin-$\frac{1}{2}$ particles at the mobility edge. Scaling
properties of multifractal measure are encoded in the $f(\alpha)$
spectrum. Here we use the direct method of Chhabra and Jensen
\cite{Chhabra89,Janssen94} to calculate this function. By using the
box probabilities $p(l)=\int_{\Omega(l)}d^2\bm{r}|\Psi(\bm{r})|^2$,
with $\Omega(l)$ being a box of linear dimension $l$, and
one-parameter families of normalized measures
$\mu_i(q,l)=p^{q}_i(l)/\sum_i p^{q}_i(l)$ for each value of $q$,
corresponding values of $\alpha(q)$ and $\tilde{f}(q)=f[\alpha(q)]$
can be calculated from
\begin{equation}\label{falpha}
    \alpha(q)=\frac{\sum_i\mu_i \ln p_i}{\ln(l/L)}, \ \ \tilde{f}(q)=\frac{\sum_i\mu_i \ln
    \mu_i}{\ln(l/L)}.
\end{equation}
In practice, linear fit to numerators vs $\ln(l/L)$ can be used to
calculate $\alpha$ and $\tilde{f}$ as the slope of fitted line. This
method requires large system sizes to avoid finite-size effect.

\begin{figure}[h]
\epsfxsize7.5truecm \epsffile{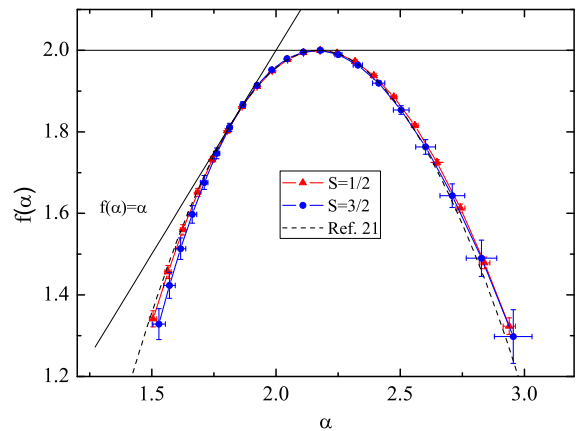} \caption{(Color online)
Singularity spectrum of critical WFs of spin-$\frac{1}{2}$ and
spin-$\frac{3}{2}$ particles with lattice sizes $1500^2$ and
$800^2$, respectively. Line $f(\alpha)=\alpha$ is tangent to
$f(\alpha)$ curve as it should be. Horizontal line shows maximum of
$f(\alpha)$ which is dimension of support ($d=2$). The data from
Ref. \onlinecite{Obuse} are included for comparison.}\label{f(a)}
\end{figure}

These exponents are related to \textit{correlation dimension}
$\tau(q)$ by a Legendre transform $\alpha(q)=d\tau(q)/dq$,
$f[\alpha(q)]=\alpha(q)q-\tau(q)$ and \textit{generalized dimension}
$D(q)$ is defined through $\tau(q)=(q-1)D(q)$.

The singularity spectrum for single critical states of
spin-$\frac{1}{2}$ and spin-$\frac{3}{2}$ particles are shown in
Fig. \ref{falpha}. Corresponding lattice sizes are $1500^2$ and
$800^2$, respectively. The eigenstates are obtained via Lanczos
algorithm. Parameters of former is picked up from phase diagram
obtained in Ref. \onlinecite{Asada02} $(E=1,W=5.952)$ and the
spin-$\frac{3}{2}$ case has $(E=3.362, W=1)$. Even thought two
states are distant in parameter space, their singularity spectrum
are the same within the error bars. Especially we obtain (for WFs in
Fig. \ref{falpha}) $\alpha_0^{1/2}=2.174\pm0.005$ and
$D^{1/2}(2)=1.66\pm0.04$. These exponents and also whole spectrum
are compatible with previous calculations based on inverse
participation ratio analysis. \cite{Mildenberger,Obuse} For
spin-$\frac{3}{2}$ case $\alpha_0^{3/2}=2.164\pm0.009$ and
$D^{3/2}(2)=1.71\pm0.05$, where $\alpha_0=\alpha(0)$. Error bars are
standard deviation of slopes in linear fitting (Fig. \ref{f(a)}).

It should be noted that, obtained results depend weakly on
realization of disorder so ensemble averaging will not change them
significantly.
\section{Summary and future work}
Localization of particles with spin $1$ and $\frac{3}{2}$ in the
presence of spin-orbit interaction is studied numerically in the
SU(2) model. In summary, spin-1 particle belongs to orthogonal
symmetry class and is always localized. Spin-$\frac{3}{2}$ case
exhibits a transition. Implication of finite-size scaling results
and multifractal analysis on critical WFs of this model is that the
transition belongs to the conventional universality in symplectic
class.

An interesting direction of future work would be investigation of
spin dependence of localization length (see Fig. \ref{ratio}) within
the Dorokhov-Mello-Pereyra-Kumar formalism for disordered wires.

{ACKNOWLEDGMENTS}

It is pleasure to thank P. Markos for helpful discussions. We would
like also to thank H. Obuse and his coworkers for sharing their data
from Ref. \onlinecite{Obuse}, R. Nourafkan for providing his Lanczos
code and F. Joibari for reading the manuscript.

\end{document}